\documentclass{elsart41}
\usepackage{graphics}
\usepackage{graphicx}
\usepackage{amssymb}
\begin{document}

\begin{frontmatter}



\title{Thermal conductivity of $R_2$CuO$_4$, with $R$ = La, Pr and Gd}
%

\author[AA]{K. Berggold\corauthref{Name1}},
\ead{berggold@ph2.uni-koeln.de}
\author[AA]{T. Lorenz},\,
\author[AA]{J. Baier},\,
\author[AA]{M. Kriener},\,
\author[AA]{D. Senff},\,
\author[BB]{S. Barilo}
\author[AA]{and A. Freimuth}\,
\address[AA]{II. Physikalisches Institut, Universit\"{a}t zu K\"{o}ln, 50937 K\"{o}ln, Germany}  
\address[BB]{Institute of Solid State $\&$ Semiconductor Physics, Belarussian Academy of Sciences, Minsk 220072, Belarus}

\corauth[Name1]{Corresponding author. Tel: +49-221-470-2911 
fax: +49-221-470-6708}

\begin{abstract}
We present measurements of the in-plane ($\kappa_{ab}$) and out-of-plane ($\kappa_{c}$) thermal conductivity of Pr$_2$CuO$_4$ and Gd$_2$CuO$_4$ single crystals. The anisotropy gives strong evidence for a large contribution of magnetic excitations to $\kappa_{ab}$, i.e. for a heat current within the CuO$_2$ planes. However, the absolute  values of $\kappa_{\rm mag}$ are lower than previous results on La$_2$CuO$_4$. These differences probably arise from deviations from the nominal oxygen stoichiometry.  This has a drastic influence on $\kappa_{\rm mag}$, which is shown by an  investigation of a  La$_2$CuO$_{4+\delta}$ polycrystal.
\end{abstract}

\end{frontmatter}

One peculiarity of low-dimensional spin systems is the possibility of large contributions to the heat transport by magnetic excitations. This has been found in 1D systems like spin ladders,\cite{sologubenko00a} whereas the situation in 2D is less clear. 
The CuO$_2$ planes in  La$_2$CuO$_4$ are a good realization of a 2D antiferromagnetic Heisenberg square lattice with a large exchange constant $J\approx 1400$\,K.\cite{kastner98a}
The thermal conductivity of La$_2$CuO$_4$ shows an unusual behavior.\cite{nakamura91a} For a heat current along the $c$ direction one low-temperature maximum is present, as it is expected for an insulator. The heat is carried by acoustic phonons, and for high temperatures Umklapp scattering yields an approximate $1/T$ behavior of $\kappa_c$.  However, for a heat current parallel to the CuO$_2$ planes  a second broad maximum arises in addition to the low-temperature maximum. This can be interpreted in terms of an additional contribution to the heat transport by magnetic excitations. Depending on the crystals, values of $\kappa_{\rm mag}^{(300\,{\rm K})}\approx15\dots 24$\,W/Km are reported.\cite{nakamura91a,yan03a,hess03a} However, a double peak of $\kappa$ could also arise from a suppression acting in a certain temperature window.\cite{hofmann01a} This requires  an additional scattering mechanism, e.g. due to a structural instability, which is present in  La$_2$CuO$_4$.\cite{braden94b} Such a structural instability is not present in the nearly iso-structural compound Sr$_2$CuO$_2$Cl$_2$, but still there is a second maximum  of $\kappa_{ab}$.\cite{hofmann03a} This gives evidence for a magnetic origin of the second maximum and 
\begin{figure}[!ht]
\begin{center}
\includegraphics[width=0.45\textwidth]{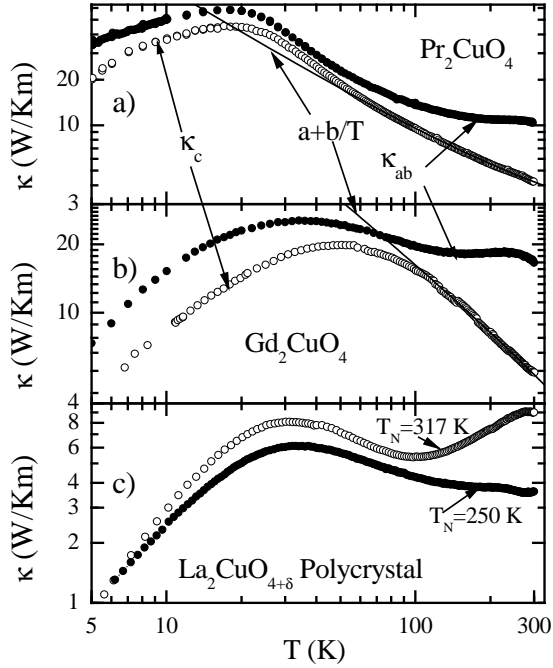}
\end{center}
\caption{Panel a) and b): $\kappa_{ab}$ and $\kappa_c$ for Pr$_2$CuO$_4$ and Gd$_2$CuO$_4$. The phononic behavior of the out-of-plane measurements at high temperatures is confirmed with $1/T$ fits (lines). Panel c): $\kappa$ of a La$_2$CuO$_{4+\delta}$ polycrystal, see text.}
\label{fig1}
\end{figure}
suggests  that a magnetic contribution to $\kappa_{ab}$ could be a common feature of all single-layered insulating cuprates. In order to investigate this, we studied $\kappa$ of  {\it R}$_2$CuO$_4$, with {\it R} = Pr, Gd. Both compounds have a  very similar CuO$_2$ square lattice and exchange constant as La$_2$CuO$_4$.\cite{johnston97a} However, for $R$\,=\,Pr the structure is stable down to lowest temperatures, whereas for $R$\,=\,Gd a structural phase transition occurs at $685$\,K.\cite{braden94a}
\par
Fig. \ref{fig1}a shows $\kappa_{ab}$ and $\kappa_c$  for Pr$_2$CuO$_4$.  For  $\kappa_c$ we observe  one low-temperature maximum and a $1/T$ behavior (solid line) above about $70$\,K. For $\kappa_{ab}$, however, a pronounced shoulder occurs at high temperatures, which leads to an anisotropy $\kappa_{ab}/\kappa_c \approx 2$ at room temperature, whereas the heights of the low-temperature maxima differ only by $20$\,\%. For Gd$_2$CuO$_4$  (Fig. \ref{fig1}b), $\kappa_c$  can also be described well by phonons. For $\kappa_{ab}$ a second broad maximum arises at high temperatures. The anisotropy at room temperature $\kappa_{ab}/\kappa_{c}\approx3$ is even larger than for Pr$_2$CuO$_4$. The low-temperature maxima differ by about $25$\,\% in height. They are shifted in temperature, which is likely to be caused by sheet defects, which are more efficient for $\kappa_{c}$.  This anisotropic behavior, which occurs in both compounds independently from the presence of a structural phase transition, gives further evidence that a magnetic contribution to $\kappa_{ab}$ is a general feature of these cuprates. We estimate the magnetic contribution at room temperature from the difference $\kappa_{\rm mag}=\kappa_	{ab}-\kappa_c$. This assumes that the phononic anisotropy is not too big, what is reasonable, since the low-temperature maxima for the different directions differ much less than the room-temperature values. In this way, we get  $\kappa_{\rm mag}^{(300\,{\rm K})}\approx6$\,W/Km for Pr$_2$CuO$_4$ and  $\approx11$\,W/Km for Gd$_2$CuO$_4$. 
\par
The different values of  $\kappa_{\rm mag}^{(300\,{\rm K})}$ raise the question, what determines the size of $\kappa_{\rm mag}$. Because $J$ does not differ too much, it is reasonable to assume that the differences mainly arise from different scattering of the heat-carrying excitations. Possible scattering mechanisms are  magnon-magnon, magnon-phonon and magnon-hole scattering. Because of the similar magnetic subsystem for all these compounds, we do not expec that magnon-magnon scattering is very different. Whether a different magnon-phonon scattering is present, is difficult to judge at the present stage of the available data. Thus, we concentrate on the study of magnon-hole scattering. 
The antiferromagnetic order in La$_2$CuO$_{4+\delta}$ is suppressed drastically by hole doping. To investigate the increase of the magnon-hole scattering, we measured $\kappa$ of a La$_2$CuO$_{4+\delta}$ polycrystal (Fig. \ref{fig1}c), which was annealed either in oxygen ($T_N=250$\,K) or in vacuum ($T_N=317$\,K). For $T_N=317$\,K a pronounced high-temperature maximum  is present. The magnitude of this maximum is lower than observed in single crystals, which partly arises from the averaging $\kappa=1/3\kappa_c+2/3\kappa_{ab}$ in a polycrystal. Moreover, $T_N$ is still lower than the maximum value $T_N^{\rm max}\approx 325$\,K.\cite{chen91a} Weak oxygen doping is also the most probable reason for the different $\kappa_{\rm mag}^{(300\,{\rm K})}$ in single crystals.\cite{nakamura91a,yan03a,hess03a} On our Pr$_2$CuO$_4$ and Gd$_2$CuO$_4$ crystals we determined $T_N=250$\,K and $295$\,K from neutron scattering and magnetic susceptibility data, respectively. For Gd$_2$CuO$_4$ we are not aware of higher $T_N$ values, but for Pr$_2$CuO$_4$ significantly larger values up to $T_N\approx 270$\,K are reported.\cite{johnston97a} This may explain the low value of $\kappa_{\rm mag}$ for Pr$_2$CuO$_4$. However, $T_N$ is not solely determined by the oxygen content, but also depends e.g. on the magnetic anisotropy. Thus, no direct conclusion concerning the hole content can be drawn from a comparison of $T_N$ for $R_2$CuO$_4$ with different $R$.
\par
In conclusion, we have measured $\kappa_{ab}$ and $\kappa_c$ of Pr$_2$CuO$_4$ and Gd$_2$CuO$_4$. The temperature dependence of $\kappa_{ab}$ and the anisotropy $\kappa_{ab}/\kappa_{c}$ give  evidence for a magnetic contribution to $\kappa_{ab}$ and suggest that this behavior is a common feature of single-layered cuprates. The magnetic contribution is suppressed drastically by charge-carrier doping, which was shown on a La$_2$CuO$_{4+\delta}$ polycrystal.
\par  
This work  was supported by the Deutsche Forschungsgemeinschaft through SFB 608.

\end{document}